%% file: node_probes.tex
\documentclass[journal,draftclsnofoot,onecolumn,12pt]{IEEEtran}

\usepackage{amsthm,amssymb,graphicx,multirow,amsmath,color,amsfonts}
\usepackage[update,prepend]{epstopdf}
\usepackage[noadjust]{cite}
\usepackage[latin1]{inputenc}
\usepackage{tikz}
\usetikzlibrary{arrows,calc}		
\usepackage{bbm} 
\usepackage{pdfpages}
\usepackage{tabulary}
\usepackage{multirow}
\usepackage{comment}
\usepackage{lipsum}
\usepackage{float}

\usepackage{lettrine}

\theoremstyle{definition}

\usepackage{array}
\usepackage{booktabs}
\usepackage{hyperref}

\usepackage{algorithm}

\usepackage{makecell}

\include{notation}

\allowdisplaybreaks 

\begin{document}
	
\graphicspath{{./figures/}}
\title{Probabilistic Analysis of Operating Modes in Cache-Enabled Full-Duplex D2D Networks}

\author{ Mansour Naslcheraghi, Constant Wetté, and Brunilde Sansò
	
	\thanks{M. Naslcheraghi and B. Sansò are   with the Department of Electrical Engineering, Polytechnique Montréal, Montréal, QC H3T 1J4, Canada (email: \{mansour.naslcheraghi, brunilde.sanso\}@polymtl.ca)}
	\thanks{Constant Wetté is with the Department of Business Area Digital Services, Ericsson, Montréal, QC H4S 0B6, Canada.} 
}

\maketitle
\thispagestyle{empty}
\pagestyle{empty}

\maketitle

\begin{abstract}
With the extensive acquisition of various mobile applications, cellular networks are facing challenges due to exponentially growing demand for high data rate, which causes a great burden on mobile core networks and backhaul links. Cache-enabled Device-to-Device (D2D) communication, which is recognized as one of the key enablers of the fifth generation (5G) cellular network, is a promising solution to alleviate this problem. However, conventional half-duplex (HD) communication may not be sufficient to provide fast enough content delivery over D2D links in order to meet strict latency targets of emerging D2D applications. In-band full-duplex (FD), with its capability of allowing simultaneous transmission and reception, can provide more content delivery opportunities, thus resulting improved spectral efficiency and latency reduction. However, given the random nature of the cached contents in user devices and users' random requests, it is unlikely to consider all involving nodes in content exchange collaborations as a pure HD or FD network. In this paper, we aim to analyze the caching perspective of a finite network of D2D nodes in which each node is endowed with FD capability and utilize a more realistic caching policy. We model and analyze all possible operating modes for an arbitrary device, which we compute the probability of occurrence of each mode along with the Probability Mass Functions (PMFs) of nodes that are operating in all possible modes. Our analysis concretely quantize all possible outcomes that strongly depend on the random nature of the caching parameters, yielding to have an accurate insight on the caching performance and all possible outcomes of the cache-enabled FD-D2D network.
\end{abstract}


\section{Introduction}
\label{sec:intro}
\lettrine[lraise=0, nindent=0em, slope=-0.5em]{F}{orecasted} by Cisco, almost 20\% of the contents account for 80\% of the total data traffic, especially multimedia contents \cite{CiscoReport}. With the fifth generation (5G) evolution, mobile edge-caching technology has attracted widespread attention because of its potential to improve system performance and enhance user experience \cite{Mobile_Edge_Caching}. The main idea behind this technology is to use the local storage of mobile devices to store popular contents and deliver them asynchronously to nearby devices through device-to-device (D2D) communications whenever they need it \cite{Negin, D2D-Caching:Mingyue}. Follow-up research utilized this technology in the context of connected vehicles as well, namely Cellular Vehicle-to-Everything (C-V2X) communications~\cite{Cache_CV2X:Cloud-Assisted, Cache_CV2X:EC-ML, Cache_CV2X:UAV, Cache_CV2X:NetworkCoding, Cache_CV2X:ADePt, Cache_CV2X:Incentive}. In C-V2X technology, vehicles with the assistance of cellular infrastructure, collaborate to offload Vehicle-to-Infrastructure (V2I) traffic using direct Vehicle-to-Vehicle (V2V) links \cite{Cache_CV2X:Cloud-Assisted}, reduce retrieve time \cite{Cache_CV2X:EC-ML, Cache_CV2X:UAV}, improve bandwidth efficiency and enhance data service performance \cite{Cache_CV2X:NetworkCoding}, and improve overall service quality and delivery rate by proactively placing relevant data in vehicles' storage \cite{Cache_CV2X:ADePt, Cache_CV2X:Incentive}. Another technology for 5G evolution is Full-Duplex (FD) radios, which allows simultaneous transmission and reception on the same time/frequency resources \cite{FD-5G:IEEE-Mag}. FD radios can potentially double the spectrum efficiency and reduce the end-to-end delay, providing efficient Self-Interference Cancellation (SIC) scheme \cite{FD_MIT, FD_Twente} is employed. This technology is broadly used in cellular communications \cite{FD-5G:IEEE-Survey,FD-Cellular:SG-Hesham, FD-5G:MyTVT} and C-V2X networks \cite{FD-CV2X:NOMA-IoT,FD-CV2X:MIMO-NOMA,FD-CV2X:Interference_Manage,FD-CV2X:Coop_Driving,FD-CV2X:Collision-Avoidance, FD-CV2X:NOMA-Decent}. In the other hand, inspired by the caching technology and FD capability, joint utilization of FD radios and caching technology promises even more advantages in comparison with its Half-Duplex (HD) counterpart by providing more content delivery opportunities, thus improving spectral efficiency and reducing end-to-end delay. This idea, namely FD-enabled caching networks proposed in \cite{MyIET} and the follow-up papers \cite{FD-Caching:MyICT, FD-Caching:Calgary, FD-Caching:MyNoF, FD-Caching:ProactiveEvolu, FD-Caching:Tan01, FD-Caching:Tan02, FD-Caching:Vu01, FD-Caching:Vu02, FD-Caching:Vu03, FD-Caching:MyICC}, put this idea  in different context. Even though FD capability brings advantages in cache-enabled networks, however, the accurate analysis of such an advantages is challenging and will be discussed in the sequel.

\subsection{Motivations and Related Work}
\label{subsec:relatedwork}
The accurate performance analysis of the cache-enabled D2D networks requires consideration of the caching mechanism not only in the cache hit performance, but also the impact of caching performance in the communication aspects, i.e., the number of transmitters/receivers, correspondingly, the amount of interference in system as well as active links which explicitly demonstrates the number of satisfied users. More precisely, the caching mechanism determines the possible transmitting users and the user operating modes (HD/FD) for a typical user which directly impacts the number of users in each mode. This consideration leads to more challenging analysis compared to the case in which the number of transmitters and receivers are assumed to be independent from the caching mechanism due to two main reasons. First, the performance depends upon the D2D network formation, which in turn depends upon the content cached in each user as well as their stochastic demands. Given different scenarios to establish D2D communications, which are either through centralized or semi-centralized manners \cite{Mobile_Edge_Caching}, it is reasonable to say that two users will initiate direct D2D link only if at least one of them finds its desired content in the other's cache, thanks to central network that have the accurate information on the users' caches and demands. It is also possible that multiple users demand for the same content which is cached by a nearby user, resulting an opportunity to target multiple receivers by a single transmission. Second, depending upon the cached contents and the users' stochastic demands, a typical node can operate in either HD or FD mode. Even when a typical node operates in FD, it does not necessarily form the more intuitive bi-directional FD (BFD) link in which two nodes exchange data with each other. Another possibility is a three node FD (TNFD) collaboration in which an intermediate node can receive its desired content from one node and concurrently serve some other node using content stored in its cache. These challenges that appear in the analysis of cache-enabled D2D and C-V2X networks with FD capability have not yet been addressed deeply and will form the basis of our contribution. On the other hand, depending on the caching policy, formation of the stochastic D2D network can be evolved rapidly. For instance, if the desired content of a typical user can be found in more than one nearby users, there will be multiple potential transmitters for a single receiver. This happens when a more realistic caching policy is being utilized, namely stochastic caching policy, thus resulting in overlap between users' caches. Also, as the FD capability directly impacts the caching performance for the reasons discussed before, the accurate analysis to get the expected number of transmitters and receivers, or accordingly, the number of potential D2D links, by considering a more realistic caching policy becomes more challenging. This opens new line of inquiry in cache-enabled D2D networks with FD capability, which will form the basis of our further contributions. As evident from the above discussion, given FD capability and depending on caching policy, a typical user can operate in different operating modes depending upon the cached content and stochastic demands and will be discussed in more detail in the next Section.

In the existing literature, e.g., see~\cite{MyIET, FD-Caching:MyICT, FD-Caching:Calgary, FD-Caching:MyNoF, FD-Caching:ProactiveEvolu, FD-Caching:Tan01, FD-Caching:Tan02, FD-Caching:Vu01, FD-Caching:Vu02, FD-Caching:Vu03, FD-Caching:MyICC}, the focus has been on the performance analysis of an arbitrary node when it is obtaining content from a nearby node by employing an optimistic caching policy, also referred as the deterministic caching policy \cite{Negin}. In this policy, it is assumed that the most popular contents are stored at the subset of users such that device $i$ caches the $i$th most popular content, without duplication of the cached content in different devices. This implicitly means that the users are scheduled to cache a set of specific contents out of all possible choices. In \cite{MyIET, FD-Caching:MyICT, FD-Caching:Calgary}, a cache-enabled and cluster-based FD-D2D network is analyzed to investigate network throughput and download time and for the worse case scenario, it is assumed that all users inside the cluster are simultaneously making random requests to retrieve their desired contents. Stochastic Geometry (SG) based analysis is being conducted in \cite{FD-Caching:MyNoF, FD-Caching:ProactiveEvolu, FD-Caching:Tan01, FD-Caching:Tan02}, in which the distribution of transmitters, relays, and receivers are determined with the caching placement probabilities inspired by the \textit{Thinning Theory} \cite{MartinBook}, and the receiver of interest is associated to the closest transmitter in the vicinity. FD-enabled transmitters in \cite{FD-Caching:Tan01, FD-Caching:Tan02} operate as relays to relay contents from the central base station (BS), when the desired content of the receiver of interest is not stored in the nearby transmitter. In the other approach \cite{FD-Caching:Vu01, FD-Caching:Vu02, FD-Caching:Vu03}, it is assumed that the portion of transmitters operate in FD mode, while the rest of users and backhaul wireless access point operate in HD mode. Therefore, the existing works focus on a very specific cases out of all possible scenarios that could occur in a cache-enabled D2D networks in which the nodes are endowed with FD capability. Specifically, utilization of a more realistic caching policy is missing in the context of FD-enabled D2D caching networks. In this paper, we overcome these shortcomings by modeling all possible operating modes along with different caching policies and investigate their impact on the caching behavior of the system, accurately. More details about the main contributions are provided next.

\subsection{Main Contributions and Outcomes}
\label{subsec:contributions}
Thus far, the existing literature have not investigated all possible operating modes for an arbitrary node in a cache-enabled C-V2X and D2D communications. Given the random nature of the cached contents in devices as well as users' random demands, it is unlikely to consider all involving nodes in content exchange collaborations as a pure FD or HD network. Hence, depending on the random circumstances (e.g. cached contents, requests, contents' characteristics, etc.) in a cache-enabled wireless FD-D2D network, a mix of FD and HD communications is a more realistic outcome. Due to these stochastic circumstances, it is also unlikely to divide the nodes into two categories of HD and FD since the portion of each category is also random. In this paper, we conduct analysis for all possible operating modes to obtain closed-form expressions for the portion of HD and FD nodes. Further, due to dynamic nature of network, utilizing the more intuitive caching policy, namely deterministic caching policy, does not entirely reflect the caching performance since the central entities of the network have to push a specific set of the contents in users' storage with such constraints. Thereby, we aim to employ more realistic caching policy, namely stochastic caching for the content placement, which is more suitable for real-life scenarios. Thus far, there is no existing work to investigate stochastic caching policy in FD-enabled D2D network. Further details on the contributions are provided in the sequel.  

\begin{itemize}
	
	\item In this paper, we employ stochastic caching policy for the content placement strategy, in which every node caches contents randomly from the large set of contents tracked and determined by the central network. Different from \cite{MyIET, FD-Caching:MyICT, FD-Caching:Calgary, FD-Caching:MyNoF, FD-Caching:ProactiveEvolu, FD-Caching:Tan01, FD-Caching:Tan02, FD-Caching:Vu01, FD-Caching:Vu02, FD-Caching:Vu03, FD-Caching:MyICC}, in which the intuitive deterministic caching is being utilized, the employed stochastic caching policy in our work causes overlapping between the users' caches, correspondingly, bringing the opportunity to deliver the content of interest to the receiver of interest through different nearby transmitters. We then compare the caching performance of the utilized random caching with its counterpart, namely deterministic caching, in terms of the number of potential receivers and outage users.      
	
	\item We carefully list all possible operating modes when nodes have the FD capability not only to relay contents from the nearby nodes (i.e. TNFD), but also exchange data simultaneously through BFD opportunities. Different from \cite{FD-Caching:Tan01, FD-Caching:Tan02}, in which the FD nodes are modeled as the relays to relay contents from the BS to the receiver of interest, here we calculate the probability of operating in FD mode such that the resulted expression accounts for both possible modes, namely BFD and TNFD modes. Also, different from \cite{FD-Caching:MyICC}, in which the operating modes are modeled based on the intuitive caching policy, here we conduct analysis for all operating modes by considering stochastic caching policy and compare the caching performance metrics with its counterpart, i.e., deterministic caching.   

	\item Further, using the closed-form expressions of the probabilities for all operating modes, we aim to derive the Probability Mass Functions (PMFs) of the number of nodes that randomly operate in different modes. Different from \cite{FD-Caching:Vu01, FD-Caching:Vu02, FD-Caching:Vu03}, in which the density of transmitters, receivers, and relay nodes are determined by the cache placement probabilities thanks to \textit{Thinning Theory}, here we obtain PMFs not only for HD transmitters/receivers and FD transceivers, but also PMFs of the outage users that fail to retrieve contents from their nearby nodes. Moreover, different from \cite{FD-Caching:MyICC}, in which the PMF is obtained only for the transmitters that actively transmit at any given time by considering intuitive deterministic caching policy, here we precisely obtain PMFs for all modes by considering stochastic caching policy and compare the PMFs of both caching policies in terms of system key parameters.
\end{itemize}

The rest of this paper is organized as follows. In Section \ref{sec:sysmodel}, we delineate the system model and introduce the key parameters. In Section \ref{sec:ProbAnalysis}, we model and analyze the node collaboration probabilities. Theoretical and simulation results are provided in Section \ref{sec:results} and finally Section \ref{sec:conclusion} concludes the paper. In the rest of paper, bold capital and bold none-capital letters account for matrices and none-bold capital letters account for sets.

\section{System Model}
\label{sec:sysmodel}
We consider a D2D network consisting of fixed $N$ number of nodes overlaying a cellular network. All nodes are assumed to have the capability of FD.  We do not impose a constraint on the cache size of devices and we assume every device has enough storage to cache a number of contents. The summary of notation is provided in Table \ref{tab:notation}. 


\begin{table}[!ht]
	\centering
	\label{tab:notation}
	\caption{Summary of notation}
	\begin{tabular}{| c || c |}
		\hline
		\textbf{Notation} & \textbf{Description} \\ \hline
		$m$ & Total number of contents \\ \hline
		$N$ & Number of users \\ \hline
		$\text{L}$ & Set of all contents \\ \hline
		$\text{U}$ & Set of users \\ \hline
		$\text{L}_N$ & Set of contents cached in $N$ users in deterministic caching \\ \hline
		$c_\ell$ & The $\ell$-th content in the library  \\ \hline
		$\rho_\kappa$  & Popularity of the $\kappa$-th content in deterministic caching\\ \hline
		$\mu_{\nu}$  & Probability of caching content $c_\nu$ in user $u_\nu$\\ \hline
		$\gamma_r$ & Skew exponent in deterministic caching \\ \hline
		$\gamma_c$ & Skew exponent in stochastic caching \\ \hline
		$\mathbf{A}$ & Total permutations \\ \hline
		$\mathbf{M}_{i \times j}(\mathbb{N})$ & Matrix of size $i \times j$ in the field domain of $\mathbb{N}$\\ \hline
		$\delta$ & Caching policy indicator; $\delta \in \{\rm o, s\}$ \\ \hline
		$\Delta$ & User operating mode indicator\\ \hline
		$\mathcal{P}_{\Delta}^{\delta}$ & \makecell{Probability of operating in mode $\Delta$ \\ with caching policy $\delta$}\\ \hline
		$\mathcal{P}_{\rm hit}^{\delta}$ & Hitting probability in caching policy $\delta$\\ \hline
		$\mathbf{\Omega^{\nu}}$ & Block matrix associate to an arbitrary user $u_\nu$\\ \hline
		$\omega_{ij}^{\nu}$ & Elements of block matrix $\mathbf{\Omega^{\nu}}$\\ \hline
		$a_{j\nu}$ & Rows indices of block matrix $\mathbf{\Omega^{\nu}}$\\ \hline
		$a_{j\nu}$ & Column indices of block matrix $\mathbf{\Omega^{\nu}}$\\ \hline
		$r_i(\mathbf{.}) $ & Operator of taking $i^{\text{th}}$ row of a matrix\\ \hline
		$\mathcal{W}_i$ & Event of missing some contents in the permutations\\ \hline
		$\mathcal{X}_\nu$ & Event of caching content $u_\nu$ in user $u_\nu$\\ \hline
		$\mathcal{Y}_\nu$ & \makecell{Event of caching the rest \\of contents in the rest of users except $u_\nu$} \\ \hline
		$\mathcal{Z}_\nu$ & Event of requesting content $c_\nu$ cached in user $u_\nu$\\ \hline
		$\phi_{ij\nu}$ & Event of occurrence of all events $\mathcal{W}_i$, $\mathcal{X}_\nu$, $\mathcal{Y}_\nu$, and $\mathcal{Z}_\nu$\\ \hline
	\end{tabular}
\end{table}

\subsection{Caching Model}
\label{subsec:CachingModel}
We assume that the central network already tracked and stored a set of popular contents. We denote the library of popular contents of size $m$ by $ \text{L} = \{ c_\ell \: | \: \ell\in\mathbb{N}, 1\le \ell \le m \}$. Each content, i.e., $c_\ell$, has an associated popularity score, which is characterized by the user requests during the period that the central network tracked those requests. This library is being sorted in terms of the popularity score in descending order, i.e., content $c_{m-1}$ is popular than that of $c_m$. Similar to \cite{MyIET, FD-Caching:MyICT, FD-Caching:Calgary, FD-Caching:MyNoF, FD-Caching:ProactiveEvolu, FD-Caching:Tan01, FD-Caching:Tan02, FD-Caching:Vu01, FD-Caching:Vu02, FD-Caching:Vu03, FD-Caching:MyICC}, we assume that all contents in the library are tracked and stored in prior and also the popularity distribution for all contents remains constant during th delivery procedures. Each user has a unique identity defined as $u_\kappa$, which is the member of set $\text{U} = \{ u_\kappa \: | \: \kappa\in\mathbb{N}, 1\le \kappa \le N \}$. Now, to determine which contents are cached in each user, we utilize two different caching policies as described in the sequel. 

\subsubsection{Deterministic Caching}
\label{subsub:optimalcaching} 
  According to this policy, each content is associated to a single user, which means that there is no overlap between cached contents in user devices. While each user has the capability of storing multiple contents, for the sake of simplicity we assume that each user caches one content and the case of caching multiple contents is left as a promising direction for future work. Under these assumptions, user $u_\kappa$ is assumed to cache content $c_\kappa$, where $c_{\kappa}$ is different across users. For instance, users $u_1$, $u_2$, and $u_3$ store contents $c_1$, $c_2$, and $c_3$, respectively. It is clear that there is only one choice to push a subset of contents $\text{L}_N = \{ c_\ell \: | \: \ell\in\mathbb{N}, 1\le \ell \le N \}$, where $\text{L}_N \subseteq \text{L}_m$ and $N \le m$, in a set of users $\text{U}$ consisting $N$ number of users. It is noted that the elements of sets $\text{L}_N$ and $\text{U}$ are mutually ordered. Without loss of generality, we assume that there is no preference in users' identification labeling since the users are sorted randomly within the set $\text{U}$ and we will justify this assumption in the analysis. We utilize Zipf distribution, which is a special case of the Riemann Zeta function and is widely used in the existing literature \cite{D2D-Caching:Ejder,D2D-Caching:Negin,FD-Caching:MyICC,D2D-Caching:Negin}. According to Zipf distribution, the popularity of content $c_\kappa$ is equivalent to the probability of requesting content $c_\kappa$. We denote this probability by $\rho_\kappa$ and is defined by
  
\begin{equation}
\label{formula:PopularityDist-optimal}
{\rho_\kappa} = \kappa^{-\gamma_r} \left( \sum_{\eta=1}^{m} \eta^{-\gamma_r}\right) ^{-1},
\end{equation}
where parameter $\gamma_r$ is the skew exponent and characterizes the popularity distribution by controlling popularity of the contents for a given library size $m$. It is clear that $\left( \sum_{\kappa=1}^{N} \rho_\kappa = 1\right) \iff (N=m) $, otherwise $\sum_{k=1}^{N} \rho_\kappa < 1$ for $N<m$.

\subsubsection{Stochastic Caching}
\label{subsub:stochasticaching}
We consider the same library $\text{L}$ and user set $\text{U}$, as described in \ref{subsub:optimalcaching}. However, each user caches a content randomly from the library, which means that there is possibility of having overlap between users' caches. User $u_\nu$ caches content $c_{\nu}$, $\nu \in \text{L}$, and it is chosen at random based on caching distribution, which will be discussed in the sequel. Choosing a set of random contents from a the corresponding set of library is permutations with repetition and there are $m^N$ choices of subsets to assign $N$ number of contents, where $N\le m$, to $N$ users, randomly. Now, we denote $\mathbf{A} \in \mathbf{M}_{m^N \times N} (\mathbb{N})$, $1 \le a_{ij} \le m$, as the permutations matrix  with repetition. Every row of this matrix is a vector denoted by $r_i(\mathbf{A})$ with entries given by the $i^{\text{th}}$ row of $\mathbf{A}$ and is called permutation. For example, if the size of library is 3, i.e., $\text{L} = \{c_1,c_2,c_3\}$, and assuming that we have two users, i.e., $\text{U}=\{u_1, u_2\}$, here are all possible $m^{N} = 3^{2}=9$ permutations to store two contents in two users: \\
$r_1(\mathbf{A}) = [ c_1, c_2 ]$,  $r_2(\mathbf{A}) = [ c_2, c_1 ]$, $r_3(\mathbf{A}) = [c_1, c_1 ]$, $r_4(\mathbf{A}) =  [c_2, c_2 ]$, $r_5(\mathbf{A}) =[c_1, c_3]$, $r_6(\mathbf{A})= [c_3, c_1 ]$, $r_7(\mathbf{A}) =[c_3, c_3]$, $r_8(\mathbf{A})= [c_2, c_3]$, $r_9(\mathbf{A}) =[c_3, c_2]$. The probability of caching content $c_\nu$ in user $u_\nu$ is similar to the popularity distribution as in Eq. \ref{formula:PopularityDist-optimal} yet with a different skew exponent $\gamma_c$. We denote this probability as $\mu_\nu$ and is given by ${\mu_{\nu}} = {\nu}^{-\gamma_c} \left( \sum_{\eta=1}^{m} \eta^{-\gamma_c}\right) ^{-1}.$ It is also noted that the whole analysis is flexible to utilize any other caching distributions and this investigation can be an extension to this work.  


\subsection{Modeling User Operating Modes}
\label{subsec:UserOperatingModes}
\begin{figure}[!ht]
	\centering
	\includegraphics[width=0.5 \textwidth]{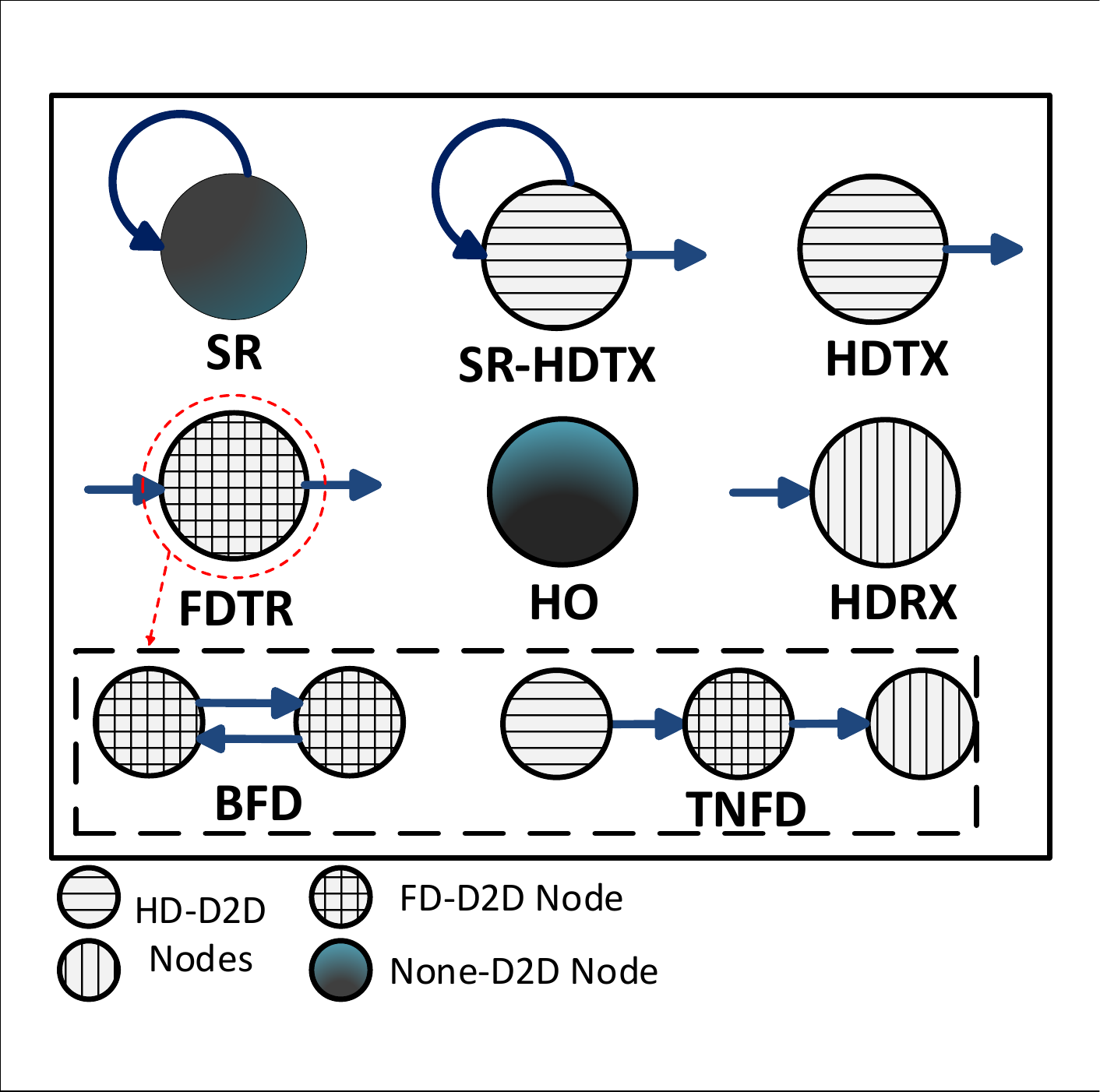}
	\caption{All possible operating modes for an arbitrary node.} 
	\label{fig:modes}
\end{figure}

Each user randomly requests a content from the library according to popularity distribution given by (\ref{formula:PopularityDist-optimal}). A pair of users can potentially initiate a D2D connection if one of them finds its desired content in the other user. Based on the information of the cached contents and users' requests, there will be different operating modes for an arbitrary user. There are six different possible operating modes for an arbitrary node as shown in Fig. \ref{fig:modes}. Definitions of the operating modes are as follows. In the sequel, $\delta \in \{ \rm d, s \}$ denotes caching policy in which ``d'' and ``s'' stand for deterministic and stochastic, respectively. 

\begin{itemize}
	\item \textbf{Self-Request (SR)}: an arbitrary user can find its desired content in its own cache. We let $\mathcal{P}_{\rm {SR}}^{\delta}$ be the probability of this mode. 
	\item \textbf{Self-Request and HD Transmission (SR-HDTX)}: an arbitrary user can find its desired content in its own cache, and can concurrently serve for other users' demand. We let $\mathcal{P}_{\rm{SR-HDTX}}^{\delta}$ be the probability of this mode.
	\item \textbf{Full-Duplex Transceiver (FDTR)}: an arbitrary user can find its desired content in its vicinity through D2D link, and can concurrently serve for other users' demand. We let $\mathcal{P}_{\rm{FDTR}}^{\delta}$ be the probability of this mode. This case can be divided into two different configurations as follows: 
	\begin{itemize}
		\item \textbf{Bi-Directional Full-Duplex (BFD):} an arbitrary user can concurrently exchange content with another user. We let $\mathcal{P}_{\rm{BFD}}^{\delta}$ be the probability of this mode.
		\item \textbf{Three-Node Full-Duplex (TNFD):} an arbitrary user can concurrently receive and transmit from and to different user. We let $\mathcal{P}_{\rm{TNFD}}^{\delta}$ be the probability of this mode.
	\end{itemize} 
	\item \textbf{Half-Duplex Transmitter (HDTX)}: an arbitrary user cannot find its desired content either in its vicinity or in its own cache, however, it can serve for other users' demand. We let $\mathcal{P}_{\rm{HDTX}}^{\delta}$ be the probability of this mode.
	\item \textbf{Half-Duplex Receiver (HDRX)}: an arbitrary user can receive its desired content via D2D link, and there is no user(s) that demand(s) for the content that is cached in this user. We let $\mathcal{P}_{\rm{HDRX}}^{\delta}$ be the probability of this mode.
	
	\item \textbf{Hitting Outage (HO)}: an arbitrary user cannot find its desired content in its vicinity or its own cache, and there is no user(s) that demand(s) for the content that is cached in this user. We let $\mathcal{P}_{\rm{HO}}^{\delta}$ be the probability of this mode. 
\end{itemize}
In the sequel, we will analyze each mode with respect to different caching policies.

\section{Analyzing Probabilities of Operating Modes and Probability Mass Functions (PMFs)}
\label{sec:ProbAnalysis}
As described in previous section, depending on the cached contents and users' requests, there are six possible operating modes for an arbitrary node. In this section, we aim to derive closed-form expressions for all operating modes by considering both deterministic and stochastic caching policies.     

\subsection{Deterministic Caching}
Given an arbitrary node and denoting $\mathcal{P}_{\Delta}^{\rm d}$ as the probability of operating mode $\Delta \in \{\rm SR, SR-HDTX, FDTR, HDTX, HDRX, HO\}$ by considering deterministic caching policy, we have the following Theorem.  
\begin{theorem}
	\label{Theorem:OptimalCaching}
	In a cache-enabled FD-D2D network endowed by deterministic caching policy, the probabilities of all possible operating modes for an arbitrary user are
	\begin{flalign}
	&\mathcal{P}_{\rm{SR}}^{\rm{d}}= \frac{1}{N} \sum_{\kappa=1}^{N} \rho_\kappa \left(1-\rho_\kappa\right)^{N-1}, \label{Formula: Probe SR} \\
	&\mathcal{P}_{\rm{SR-HDTX}}^{\rm{d}} = \frac{1}{N} \sum_{\kappa=1}^{N}\rho_\kappa \left(1-\left(1-\rho_\kappa \right)^{N-1}\right), \label{Formula: Probe SR-HDTX}\\
	&\mathcal{P}_{\rm{FDTR}}^{\rm{d}} = \frac{1}{N} \sum_{\kappa=1}^{N}\left(\mathcal{P}_{\rm{hit}}^{\textup{d}}-\rho_\kappa \right)\left(1-\left(1-\rho_\kappa \right)^{N-1}\right), \label{Formula: Probe FDTR}\\
	&\mathcal{P}_{\rm{HDRX}}^{\rm{d}} = \frac{1}{N} \sum_{\kappa=1}^{N}\left(\mathcal{P}_{\rm{hit}}^{\rm{d}}-\rho_\kappa \right)\left(1-\rho_\kappa\right)^{N-1},\label{Formula: Probe HDRX}\\
	&\mathcal{P}_{\rm{HDTX}}^{\rm{d}} = \frac{1}{N} \sum_{\kappa=1}^{N}\left(1-\mathcal{P}_{\rm{hit}}^{\textup{d}}\right)\left(1-\left(1-\rho_\kappa \right)^{N-1}\right), \label{Formula: Probe HDTX}\\
	&\mathcal{P}_{\rm{HO}}^{\rm{d}} = \frac{1}{N} \sum_{\kappa=1}^{N}\left(1-\mathcal{P}_{\rm{hit}}^{\rm{d}}\right)\left(1-\rho_\kappa \right)^{N-1}, \label{Formula: Probe HO}
	\end{flalign}
	where $\rho_\kappa$ is given in Eq. \eqref{formula:PopularityDist-optimal} and $\mathcal{P}_{\rm{hit}}^{\textup{d}}=\sum_{\kappa=1}^{N}\rho_\kappa$, which is the hitting probability at which a given random request of an arbitrary user ``hits'' the given subset of the probability space defined in subsection \ref{subsub:optimalcaching}. 
\begin{IEEEproof}
	See Appendix \ref{appendix:ModesProbe}. 
\end{IEEEproof}

In Theorem \ref{Theorem:OptimalCaching}, the expression of FDTR mode is summation of BFD and TNFD modes as demonstrated in \figref{fig:modes}, i.e.,  
\begin{equation}
\mathcal{P}_{\textup{FDTR}}^{\rm d} = \mathcal{P}_{\textup{BFD}}^{\rm d} + \mathcal{P}_{\textup{TNFD}}^{\rm d}, 
\end{equation}
where, 
	\begin{flalign}
	&\mathcal{P}_{\textup{BFD}}^{\rm d} = \frac{1}{N} \sum_{\kappa=1}^{N}\left(\mathcal{P}_{\textup{hit}}^{\rm d}-\rho_\kappa \right)\rho_\kappa, \label{Formula:ProbeBFD}\\
	&\mathcal{P}_{\textup{TNFD}}^{\rm d} = \frac{1}{N} \sum_{\kappa=1}^{N}\left(\mathcal{P}_{\textup{hit}}^{\rm d}-\rho_\kappa \right)\left(1-\rho_\kappa-\left(1-\rho_\kappa\right)^{N-1}\right). \label{Formula:ProbeTNFD} 
	\end{flalign}
\begin{IEEEproof}
	See Appendix \ref{appendix:ModesProbe}. 
\end{IEEEproof}


\subsection{Stochastic Caching}
Given an arbitrary node and denoting $\mathcal{P}_{\Delta}^{\rm s}$ as the probability of operating mode $\Delta$ by considering stochastic caching policy, we have the following Theorem.  
\begin{theorem}
	\label{Theorem:StochasticCaching}
	In a cache-enabled FD-D2D network endowed with stochastic caching policy, the probabilities of all possible operating modes for an arbitrary user are 
\begin{flalign}
&\mathcal{P}_{\rm SR}^{\rm s} =\sum_{x \in \text{L}^1} {\rho_x}\left(1-\rho_x\right)^{N-1} \mu_x, \label{Formula:RandomSR} \\
&	\mathcal{P}_{\rm SR-HDTX}^{\rm s} =\sum_{x  \in \text{L}^{1}}\rho_x\left(1-\left(1-\rho_x\right)^{N-1}\right) \mu_x, \label{Formula:RandomSR-HDTX}\\
&	\mathcal{P}_{\rm FDTR}^{\rm s} =  \mathbf{R}^{\rm T} \mathbf{C}_{{\alpha}} \left(\mathbf{e}-\mathbf{H}\right) \mathbf{C}_{\mathbf{\beta}}, \label{Formula:RandomFDTR}\\
&\mathcal{P}_{\rm HDRX}^{\rm s} =  {\left(\mathbf{e-R}\right)}^{\rm T} \mathbf{C}_{\alpha} \left(\mathbf{e} - \mathbf{H}\right)\mathbf{C}_{\mathbf{\beta}},\label{Formula:RandomHDRX}\\
&\mathcal{P}_{\rm HDTX}^{\rm s} =  \mathbf{R}^{\rm T} \mathbf{C}_{\mathbf{\alpha}} \mathbf{H} \mathbf{C}_{\beta}, \label{Formula:RandomHDTX}\\
&\mathcal{P}_{\rm HO}^{\rm s} =  \left(\mathbf{e}-\mathbf{R}\right)^{\rm T} \mathbf{C}_{\alpha} \mathbf{H} \mathbf{C}_{\beta}, \label{Formula:RandomO-D2D}
\end{flalign}
where, 
\begin{align}
		\mathbf{H} = \begin{bmatrix}
			\rm Pr \left(\mathcal{W}_1 = 0 \:|\: r_1(\mathbf{\Omega^{1}}) \right) & \rm Pr \left(\mathcal{W}_2 = 0 \:|\: r_2(\mathbf{\Omega^{1}}) \right) & \dots & \rm Pr \left(\mathcal{W}_{m^{N-1}} = 0 \:|\: r_{m^{N-1}}(\mathbf{\Omega^{1}}) \right) \\
			\vdots & \vdots & \vdots & \vdots \\ 
			\rm Pr \left(\mathcal{W}_1 = 0 \:|\: r_1(\mathbf{\Omega^{m}}) \right) & \rm Pr \left(\mathcal{W}_2 = 0 \:|\: r_2(\mathbf{\Omega^{m}}) \right) & \dots & \rm Pr \left(\mathcal{W}_{m^{N-1}} = 0 \:|\: r_{m^{N-1}}(\mathbf{\Omega^{m}}) \right)
		\end{bmatrix}_{m \times m^{N-1}}, \notag
	\end{align}
	\begin{align} 
		\mathbf{e} = \begin{bmatrix}
			1\\
			1\\
			\vdots\\
			1
		\end{bmatrix}_{m^N \times 1}, \mathbf{C}_{\alpha} = \begin{bmatrix}
			\mu_1 & 0 & \dots & 0\\
		0 & \mu_2 & \dots & 0\\
		\vdots & \vdots & \ddots & \vdots\\
		0 & 0 & \dots & \mu_m\\	
	\end{bmatrix}_{m \times m},  \mathbf{C}_{\beta} = \begin{bmatrix}
		\rm Pr \left(\mathcal{Y}_1 \:|\: r_1(\mathbf{\Omega^{1}}) \right)\\
		\vdots \\
		\rm Pr \left(\mathcal{Y}_{m^{N-1}}  \:|\: r_{m^{N-1}}(\mathbf{\Omega^{1}}) \right) \\	
	\end{bmatrix}_{m^{N-1} \times 1}, \notag
	\end{align}
\begin{align}
	\mathbf{R} = \begin{bmatrix}
		\rm Pr \left(\mathcal{Z}_1 \:|\: r_1(\mathbf{\Omega^{1}}) \right)\\
		\vdots \\
		\rm Pr \left(\mathcal{Z}_{m} \:|\: r_{1}(\mathbf{\Omega^{m}}) \right) \\	
	\end{bmatrix}_{m \times 1}, \rm Pr \left(\mathcal{W}_i = 0 \:|\: r_i(\mathbf{A}) \right) = 1 - \sum\limits_{\scriptstyle \{\theta_{j}\} \in r_i(\mathbf{A})  \hfill\atop	\scriptstyle \theta_j \neq \theta_k \hfill} {\rho_{\theta_{j}}}, \notag
\end{align} 
\begin{align}
	\rm Pr (\mathcal{Y}_\nu \:|\: r_j(\mathbf{\Omega^{\nu}})) = \prod_{\{\theta_i | \theta_i \neq \nu\} \in r_j(\mathbf{\Omega^{\nu}})}\mu_{\theta_i}, 	\quad \rm Pr (\mathcal{Z}_\nu \:|\: r_j(\mathbf{\Omega^{\nu}})) = 1-(1-\rho_\nu)^{N-1}, \quad \mathbf{\Omega^{\nu}} = \mathbf{A} (a_{j\nu}; b_k ), \notag
\end{align}
\begin{align}
	a_{j\nu}= \{ (i+ {\mathbbm{1}_\nu} (\nu-1))m^{N-1} \:|\: i \in \{1,2,\dots,m\} \}, \quad b_k=\{1,2,\dots,N\}, {\mathbbm{1}_\nu } = \left\{ \begin{array}{l}
			\begin{array}{*{20}{c}}
				0&{;\nu  = 1}
			\end{array}\\
			\begin{array}{*{20}{c}}
				1&{; o.w.}
			\end{array}
		\end{array}\right. \notag
\end{align}


%

%
%

%
%

\end{theorem}

\begin{IEEEproof}
	See Appendix \ref{proof for random caching theorem}. 
\end{IEEEproof}
Using the above results, we can obtain the probability that an arbitrary node operates in the transmitting mode and receiving mode, which are given by the following Corollary. These probability will be used in the next sections. 
\begin{cor}
	\label{Corollary:TX/RX Probes}
	While the detailed analysis of each operating mode gives an insight on the caching performance with respect to all possibilities, it is also interesting to quantize the caching performance by considering more general operating modes such as transmitter/receiver and FD/HD modes. the new metrics of interest, namely the probability of operating in transmitter mode denoted by  $\mathcal{P}_{\rm TX}^{\delta}$,  receiver mode denoted by $\mathcal{P}_{\rm RX}^{\delta}$, FD mode denoted by $\mathcal{P}_{\rm FD}^{\delta}$, and HD mode denoted by $\mathcal{P}_{\rm HD}^{\delta}$ are defined as follows:
	\begin{flalign}
		& \mathcal{P}_{\rm HD}^{\delta} =\mathcal{P}_{\rm SR-HDTX}^{\delta} + \mathcal{P}_{\rm HDRX}^{\delta} + \mathcal{P}_{\rm HDTX}^{\delta}, \label{Formula:PHD}\\
		& \mathcal{P}_{\rm FD}^{\delta} = \mathcal{P}_{\rm FDTR}^{\delta}, \label{Formula:PFD}\\
		& \mathcal{P}_{\rm TX}^{\delta} = \mathcal{P}_{\rm SR-HDTX}^{\delta} + \mathcal{P}_{\rm HDTX}^{\delta} + \mathcal{P}_{\rm FDTR}^{\delta}, \label{Formula:PTX}\\
		& \mathcal{P}_{\rm RX}^{\delta} = \mathcal{P}_{\rm FDTR}^{\delta} + \mathcal{P}_{\rm HDRX}^{\delta}. \label{Formula:PRX}
	\end{flalign} 
\end{cor}
\begin{IEEEproof}
	A transmitting node should operate either in SR-HDTX, HDTX, or FDTR mode, which means $	\mathcal{P}_\textup{TX}^{\delta} = \mathcal{P}_\textup{SR-HDTX}^{\delta} + \mathcal{P}_\textup{HDTX}^{\delta} + \mathcal{P}_{\rm FDTR}^{\delta}.$ Similar logic applies for the rest.
\end{IEEEproof}
Also, one can say: \\ 	$\mathcal{P}_{\rm SR}^{\delta} + \mathcal{P}_{\rm SR-HDTX}^{\delta} + \mathcal{P}_{\rm FDTR}^{\delta} + \mathcal{P}_{\rm HDTX}^{\delta} + \mathcal{P}_{\rm HDRX}^{\delta} + \mathcal{P}_{\rm HO}^{\delta} = 1$.   
\end{theorem}

\subsection {PMFs}
\label{subsec:PMFs}
Here, we are interested to obtain the PMFs of each operating mode. We have the following Theorem.   
\begin{theorem}
	The number of nodes operating in each mode denoted by $X_\Delta$ is a Binomial random variable, i.e., $X_\Delta \sim B(N,\mathcal{P}_\Delta^\delta)$, and its probability mass function (PMF) is
	\begin{align}
f(N, n_\Delta, \mathcal{P}_{\Delta}^{\delta}) & = \Pr (X_\Delta = n_\Delta) \notag \\& = \bigg(\begin{array}{c}N\\ n_\Delta\end{array}\bigg)(\mathcal{P}_{\Delta}^{\delta})^{n_\Delta}\left(1-\mathcal{P}_{\Delta}^{\delta}\right)^{N-n_\Delta},
	\end{align}
where $\left( \begin{array}{l}
	N\\
	n_\Delta
\end{array} \right) = \frac{N!}{(N-n_\Delta)! n_\Delta !}$. 
\end{theorem}

\begin{IEEEproof}
The probability of getting exactly $n_\Delta$ (interpreted as the number of nodes operating in mode $\Delta$) successes in $N$ independent Bernoulli trials gives leads to Binomial distribution. In fact, $n_\Delta$ successes occur with probability $\mathcal{P}_\Delta^\delta$ and $N-n_\Delta$ failures occur with probability $(1-\mathcal{P}_\Delta^\delta)^{N-n_\Delta}$. However, it is possible to have $n_\Delta$ successes out of $N$ trials, in which there are $\bigg(\begin{array}{c}N\\n_\Delta\end{array}\bigg)$ different ways of distributing $n_\Delta$ in a sequence of $N$ trials.
\end{IEEEproof}

\section{Analytical/Simulation Results and Discussions}
\label{sec:results}
\begin{figure}[!ht]
	\centering
	\includegraphics[width=0.75 \textwidth]{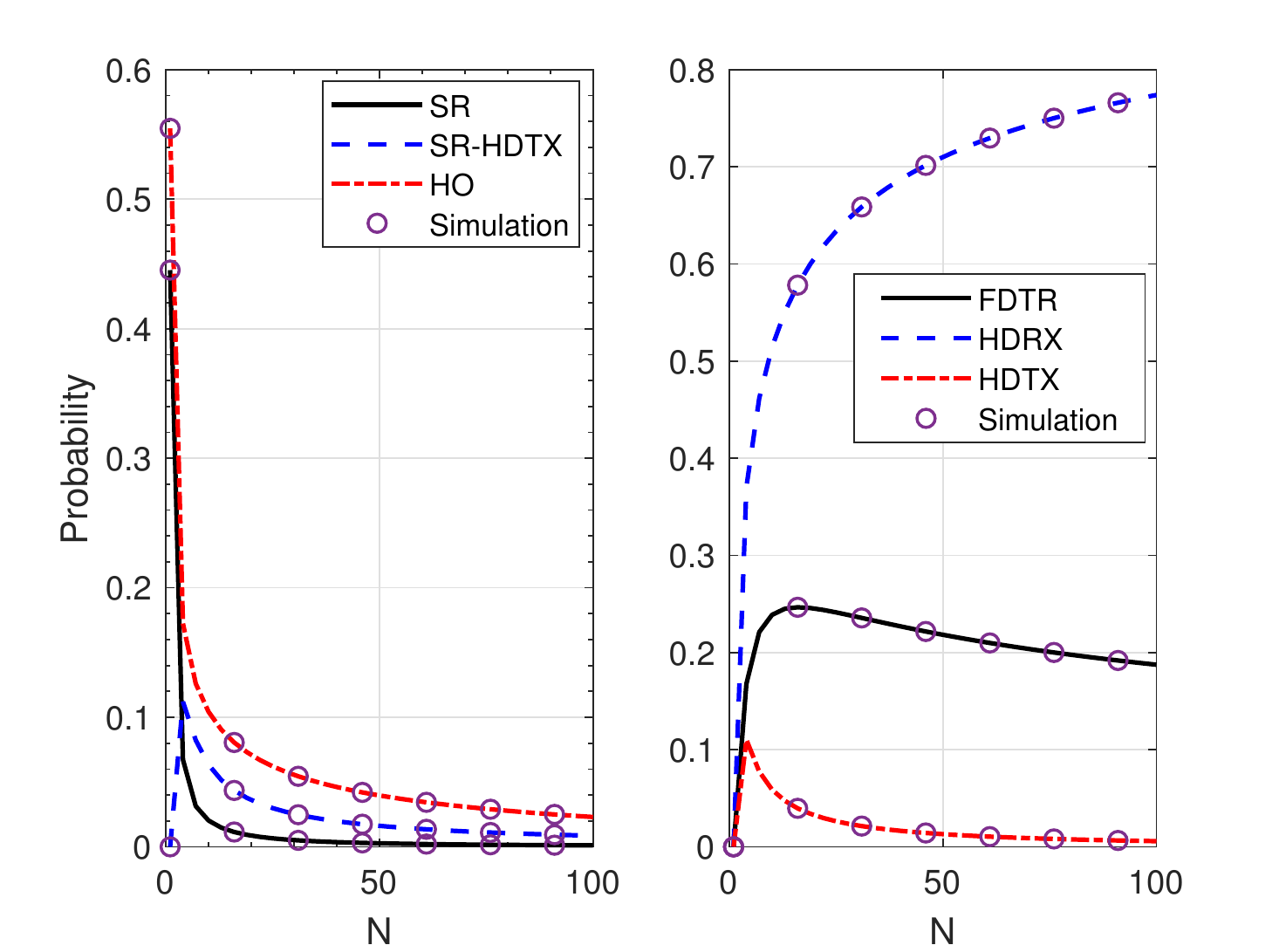}
	\caption{Deterministic caching with $M=500$, $\gamma_r = 0.8$} 
	\label{fig:opt_prob_subplot}
\end{figure}

\begin{figure}[!ht]
	\centering
	\includegraphics[width=0.75 \textwidth]{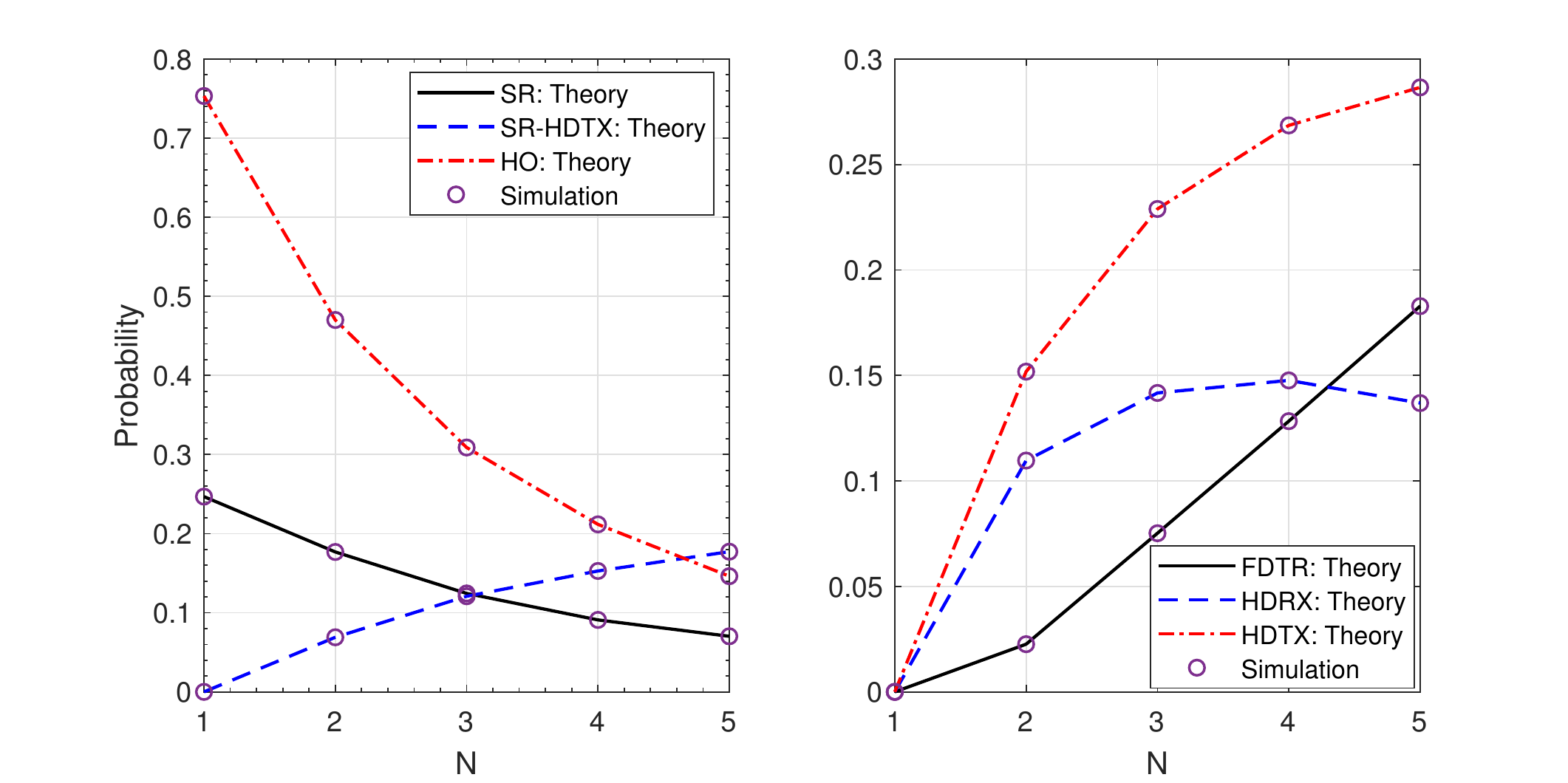}
	\caption{Limited results of the stochastic caching with $M=7$, $\gamma_r = 0.8$, $\gamma_c = 1.6$} 
	\label{fig:ran_limited_subplot}
\end{figure}

\begin{figure}[!ht]
	\centering
	\includegraphics[width=0.75 \textwidth]{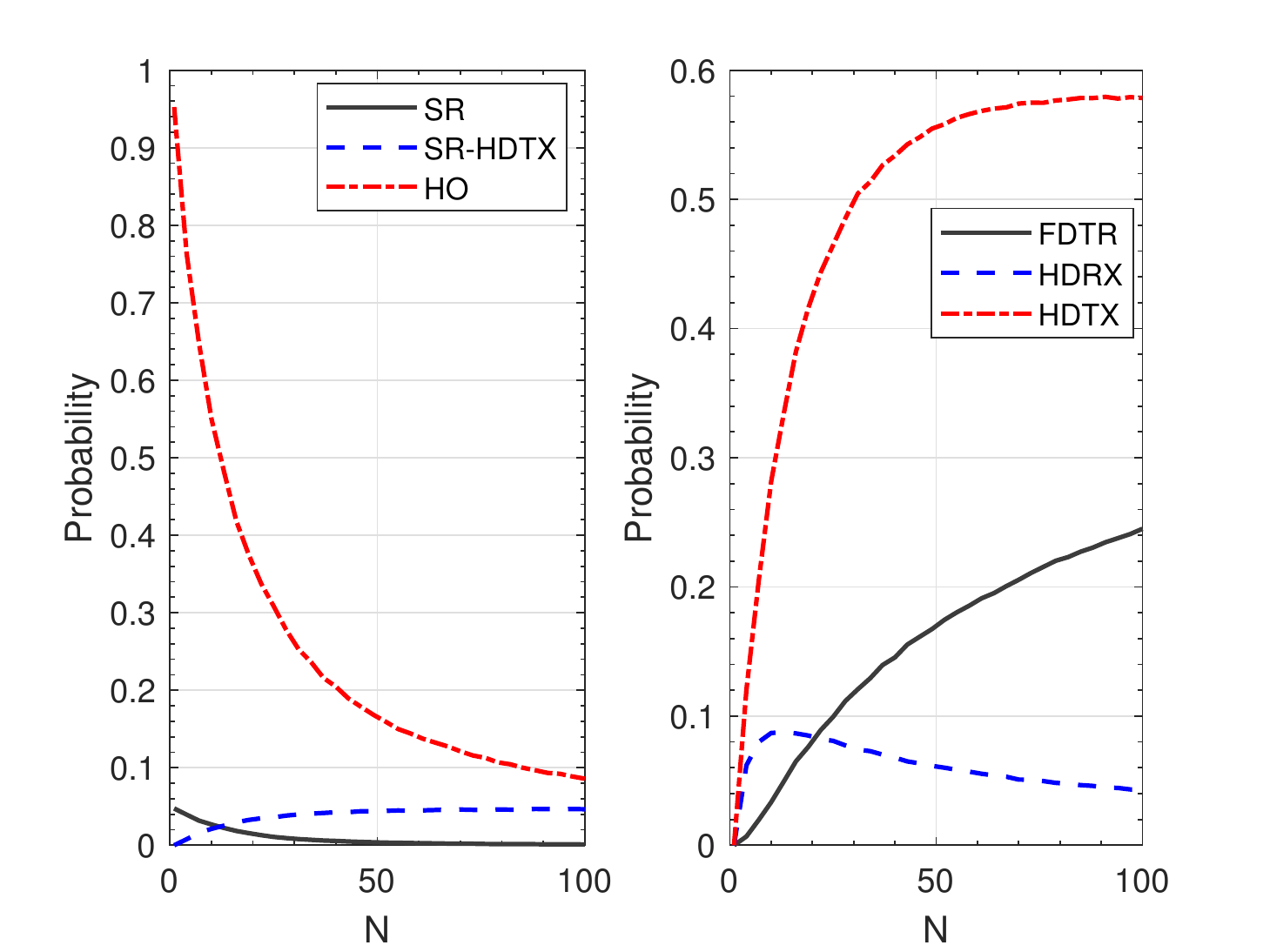}
	\caption{Stochastic caching with $M=500$, $\gamma_r = 0.8$, $\gamma_c = 1.6$} 
	\label{fig:ran_prob_subplot}
\end{figure}

\begin{figure}[!h]
\centering
	\includegraphics[width=1.1\textwidth]{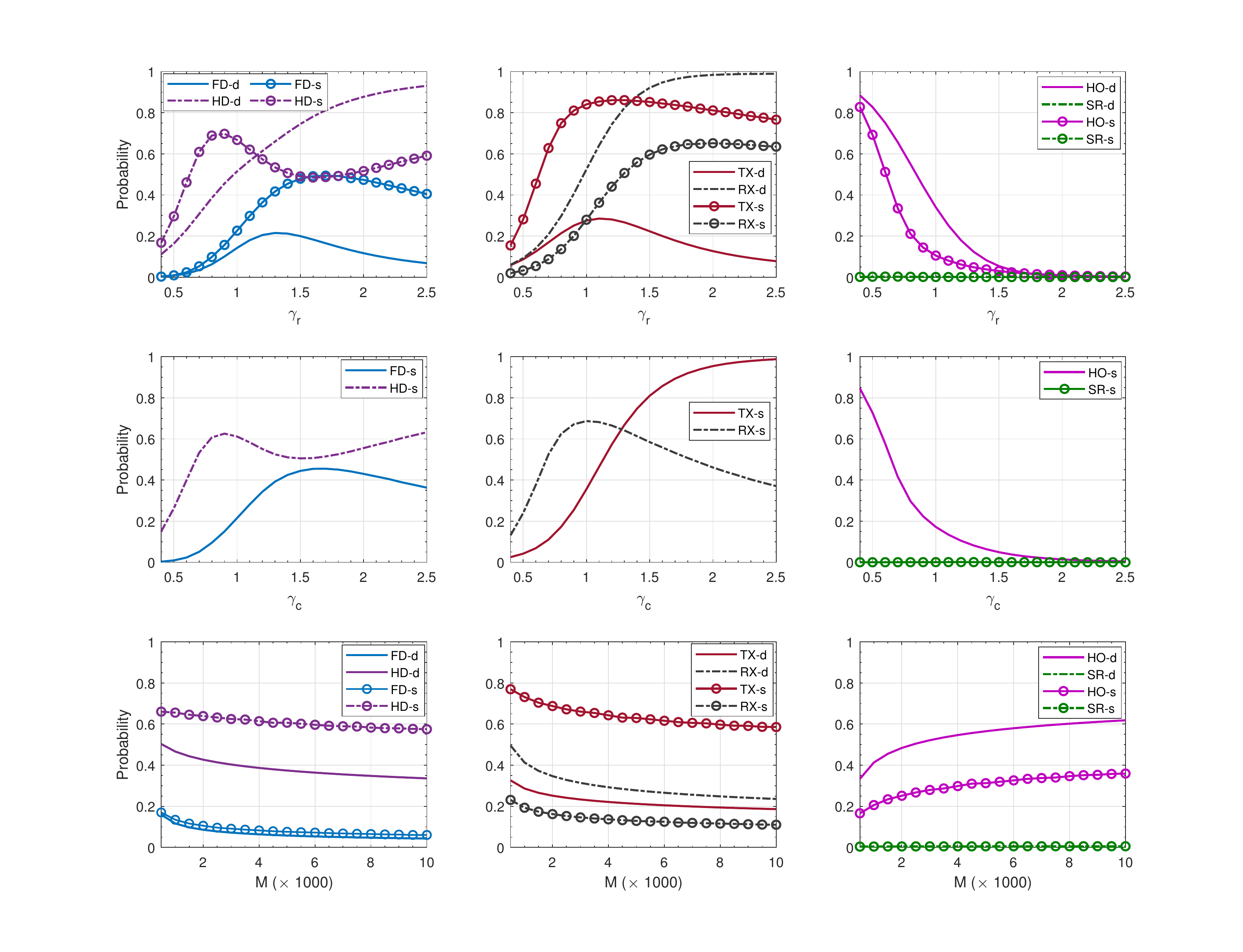}
	\caption{Comparison of probabilities versus $\gamma_r$ , $\gamma_c$, and $M$.} 
	\label{fig:optran_comp}
\end{figure}

\begin{figure}[!h]
	\centering
	\includegraphics[width=1.1 \textwidth]{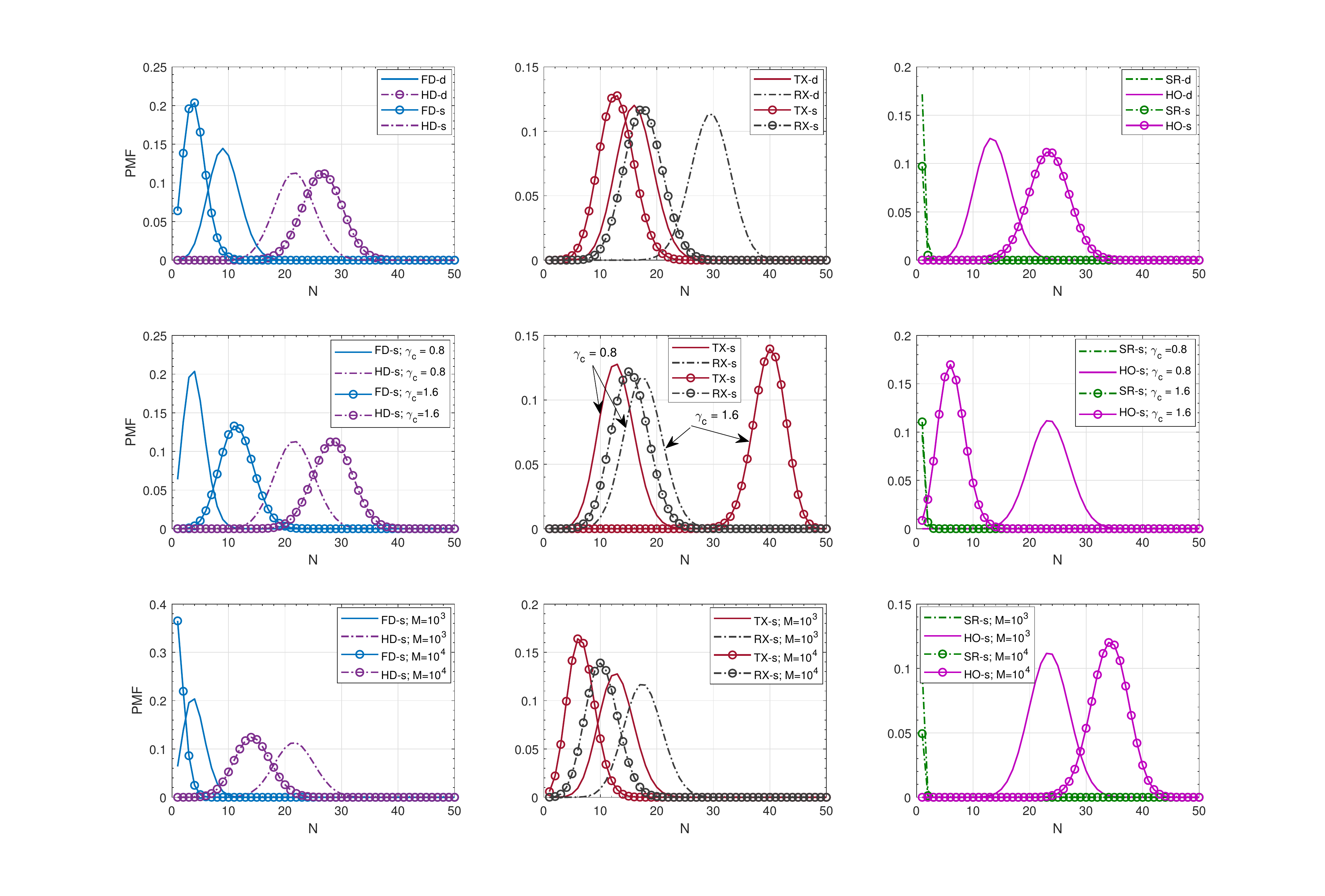}
	\caption{Comparison of PMFs with respect to $\gamma_c$ and $M$.} 
	\label{fig:optran_pmfs_comp}
\end{figure}

Fig. \ref{fig:opt_prob_subplot} illustrates the probability of each operating modes versus the number of user devices $N$ when \textit{Deterministic Caching} policy is being employed. These results are delineated for a typical value of the skew exponent $\gamma_r = 0.8$ while a broad range of values are being utilized in the further results and we discuss them later. As can be observed Fig. \ref{fig:opt_prob_subplot}, users are less likely to collaborate in SR, SR-HDTX, and HO modes. By increasing the number of user devices, the probabilities of these events tend to almost zero. The highest chance for these three events occur when $N$ is very low due two main reasons: i) the lower number of users leads to have a less diversity in available cached contents, correspondingly, the highest rank of the contents are of more interest to a very low number of users, and ii) the very low number of users eliminates the chance to form other operating modes since there are a few users in the vicinity. However, in contrast, the chance of forming operating mode HDRX is increasing since there is more diversity in the cached contents, while the chance of FDTR and HDTX decay after reaching the maximum peak for the values of $N$ less than 50. As can be seen, HDRX mode is the dominating operating mode among all modes when $N$ increases and users are more likely to collaborate in HDRX mode. The reason is explicitly because of FD capability that provides more potential delivery opportunities. \figref{fig:ran_limited_subplot} and \figref{fig:ran_prob_subplot} demonstrate the results when the \textit{Stochastic Caching} policy is being employed with a typical values of $M$, $\gamma_c$, and $\gamma_r$, yet the results with a broad range of values is conducted in further results and we will discuss them, later. The difference between these two figures is the number of users $N$. Due to high complexity order of Eqs.  \eqref{Formula:RandomSR}, \eqref{Formula:RandomSR-HDTX} \eqref{Formula:RandomFDTR}, \eqref{Formula:RandomHDTX}, \eqref{Formula:RandomHDRX}, and \eqref{Formula:RandomO-D2D},  \figref{fig:ran_limited_subplot} is delineated with a limited values of $N$ and $M$, while Fig. \ref{fig:ran_prob_subplot} demonstrates the simulation results with higher values of $N$ and $M$. While the overall behavior of the \textit{Stochastic Caching} policy is similar to \textit{Deterministic Caching} yet there are key differences between them as follows. First, The HO mode is more accented in the \textit{Stochastic Caching}, which means that in the realistic caching policy, namely \textit{Stochastic Caching}, the outage probability is higher than that of the case when \textit{Deterministic Caching} policy is being employed. Another key difference is the major difference between both caching policies with respect to FDTR mode, in which the FDTR mode is always increasing when $N$ increases. The main reason behind this behavior comes from the high chance of overlapping in the cached contents when \textit{Stochastic Caching} policy is being employed. In the other words, because of the overlapping, in one hand, there is a high chance of finding users' desired content in their vicinity, and in the other hand, FD provides more higher potential delivery opportunities. However, in \textit{Deterministic Caching}, there is no overlap. 
\figref{fig:optran_comp} demonstrates the comparison of the metrics in Eqs.\eqref{Formula:PHD}, \eqref{Formula:PFD}, \eqref{Formula:PTX}, \eqref{Formula:PRX} with respect to both caching policies by considering a broad range of key parameters. This figure includes i) comparison of FD and HD modes in both caching policies labeled by HD-$\delta$ and FD-$\delta$ (figures in the first column), ii) comparison of operating in transmitter and receiver modes labeled by TX-$\delta$ and RX-$\delta$,  respectively (figures in the second column), and iii) comparison of SR and HO modes in both caching policies  labeled by SR-$\delta$ and HO-$\delta$ (figures in the third column). Simulation parameters utilized for this figure are summarized in Table \ref{tab:simulationparams}.     
\begin{table}[!ht]
	\centering
	\caption{Simulation Parameters in \figref{fig:optran_comp}}
\begin{tabular}{c|ccc}
	$\mathcal{P}_{\Delta}^{\delta}$ vs. & \multicolumn{3}{c}{Parameters} \\ \toprule
	& N        & M         & $\gamma_c$       \\ \hline
	$\gamma_r$  & 100         & $10^{4}$          &   1.6       \\ \midrule
	& N        & M         & $\gamma_r$       \\ \hline
	$\gamma_c$  &  100        &   $10^{4}$         &   2.5       \\ \midrule
	& N        & $\gamma_r$        & $\gamma_c$       \\ \hline
	M   &  50        &  0.8         &  1.6       
\end{tabular}
\label{tab:simulationparams}
\end{table}
According to \figref{fig:optran_comp}, the probability of HD mode is higher than that of the FD mode with respect to all key parameters $\gamma_r$, $\gamma_c$, and $M$. The higher values of $\gamma_r$ and $\gamma_c$ demonstrates the higher redundancy in requests and cached contents, respectively. In the other words, the higher is the $\gamma_r$, the higher is the chance of requesting high ranked contents, and similarly, the higher is the $\gamma_c$, the higher is the chance of caching high ranked contents. As for the size of the library $M$, the higher is the $M$, the higher is the diversity in the popularity of the contents. As can be observed, increasing $M$ has a minor impact on the probabilities of the operating modes in comparison with the impact of parameters $\gamma_r$ and $\gamma_c$. Even though HD mode demonstrate higher probability in those cases, however, it is because of the explicitly key role of the FD capability, especially TNFD mode, that provides more opportunity for potential content deliveries. Similar explanations are valid for the TX and RX modes. We can find the optimal values of $\gamma_c$ that maximizes the probabilities of interest, by solving the following problem: 
\begin{flalign}
	\frac{\partial \mathcal{P}_{\Delta^\prime}^{s}}{\partial \gamma_c}=0, \label{Formula:optimal_zc}
\end{flalign}
where $\Delta^\prime \in \{ \rm HD, FD, TX, RX \}$. In this paper, we solved the above equation through the simulations and obtained $\gamma_c \approx 0.8$ for HD and RX modes, and $\gamma_c \approx 1.8$ for FD mode. It is also noted that there is no optimal value of $\gamma_c$ for TX mode as can be observed from \figref{fig:optran_comp}. As for the HO mode, higher redundancy, namely higher values of $\gamma_r$ and $\gamma_c$, demonstrates lower chance of outage for the \textit{Stochastic Caching} policy in comparison with the \textit{Deterministic Caching} policy. This is because of the overlapping as discussed in Figs. \ref{fig:opt_prob_subplot}, \ref{fig:ran_limited_subplot}, and \ref{fig:ran_prob_subplot}. Also, similar explanations of the Figs. \ref{fig:opt_prob_subplot}, \ref{fig:ran_limited_subplot}, and \ref{fig:ran_prob_subplot} apply to SR mode.    
\figref{fig:optran_pmfs_comp} illustrates the PMF of the operating modes. For the sake of visibility and presentation, all results are shown in a continues manner while the original ones are discrete. In this figure, the set of subfigures in the first row demonstrates the comparison between \textit{Deterministic Caching} and \textit{Stochastic Caching} policies, while the second and third rows illustrates the PMFs in the \textit{Stochastic Caching} with respect to skew exponent $\gamma_c$ and library size $M$, respectively. As can be seen from this figure, the expected number of HD users is more than that of FD users for all cases. Also, the expected number of users operating in \textit{Deterministic Caching} policy is higher than that of FD ones; the reason can be interpreted from the results in \figref{fig:opt_prob_subplot} and \figref{fig:ran_prob_subplot} since the probability of HD mode is higher than FD modes and the same explanations of those figures are valid here. Further, higher redundancy in the cached contents increases the number of collaborating HD/FD users, i.e., Caching performance with $\gamma_c = 1.6$ provides more collaborating nodes in comparison with the case when $\gamma_c = 0.8$. As for the TX/RX modes, the role of the skew exponent $\gamma_c$ is the same as the role in HD/FD modes. Further, increasing the library size $M$ yields reduction in the number of nodes operating in HD/FD and TX/RX modes, conversely, increases the number of HO and SR modes.          

\section{Concluding Remarks}
\label{sec:conclusion}
In this paper, we have utilized deterministic and stochastic caching policies in cache-enabled FD-D2D network and derived closed form expressions for the probabilities of different operating modes that appear in the network. We used these closed-form expressions to obtain the Probability Mass Functions (PMFs) of the number of nodes that operate in each mode, which ultimately facilitated caching performance analysis in terms of the key parameters such as skew exponents, library size, and number of users. Exploring real world dataset gathered from the operators and mapping them into the obtained results, will be a useful extension to this work. Moreover, employing alternative tools such as \textit{Graph Theory} to simplify the complexity of closed-form expressions in the stochastic caching and reduce the computational time, would be another worthwhile work.

\appendix

\subsection{Proof for Theorem \ref{Theorem:OptimalCaching}} 
\label{appendix:ModesProbe}
From \figref{fig:modes}, we can infer that the probability of occurrence of each mode at an arbitrary user $u_\kappa$ depends on two different events: i) request of the user $u_\kappa$ (we denote this event by $\mathcal{V}$ and the probability of this event by $\mathcal{P}_{\Delta}^{v,\kappa
}$) and ii) requests from other users for the content cached in user $u_\kappa$. We denote this event by $\mathcal{W}$ and the probability of this event by $\mathcal{P}_{\Delta}^{w,\kappa}$. The joint probability of both events, i.e., $\mathbb{P} (\mathcal{V},\mathcal{W})$ gives the probability of the operating mode $\Delta$ for a specific node $u_\kappa$. Since the requests at all users are independent from each other, we can say that $\mathbb{P} (\mathcal{V},\mathcal{W}) = \mathbb{P}(\mathcal{V})\mathbb{P}(\mathcal{W})=\mathcal{P}_{\Delta}^{v,\kappa
}\mathcal{P}_{\Delta}^{w,\kappa
}$. Now, by using the \textit{law of total probability} and given the deterministic caching policy, the probability of operating mode $\Delta$ denoted by $\mathcal{P}_\Delta^{\rm d}$ for an arbitrary node can be defined by 
\begin{equation}
\label{P vw}
\mathcal{P}_{\Delta}^{\rm d} = \sum_{\kappa=1}^{N} \mathcal{P}_{\Delta}^{v,\kappa} \mathcal{P}_{\Delta}^{w,\kappa} \mathcal{P}_{u_\kappa},
\end{equation}
where, $\mathcal{P}_{u_\kappa}$ is the probability of choosing an arbitrary user among $N$ users uniformly at random. Due to lack of space, we provide the proof for the FDTR mode, however, the approach remains the same for the other modes as well. Now, let us define two binary random variables $\mathcal{I}_\kappa$ and $\mathcal{J}_{\tau,\kappa}$ for $u_\kappa$ as follows.
\begin{equation}
{\mathcal{I}_\kappa} = \left\{ \begin{array}{l}
\begin{array}{*{20}{c}}
0&{;{ u_\kappa \textup{ cannot find its desired content}}}
\end{array}\\
\begin{array}{*{20}{c}}
1&{;{ u_\kappa \textup{ can find its desired content,}}}
\end{array}
\end{array} \right.
\end{equation}
\begin{equation}
{\mathcal{J}_{\tau,\kappa}} = \left\{ \begin{array}{l}
\begin{array}{*{20}{c}}
0&{;{ u_\tau \textup{ does not demands for content $c_\kappa$}}}
\end{array}\\
\begin{array}{*{20}{c}}
1&{;{ u_\tau \textup{ demands for content $c_\kappa$.}}}
\end{array}
\end{array} \right.
\end{equation}
The probability $\Pr \left(\mathcal{I}_\kappa=1\right)$ is equivalent to the situation that $u_\kappa$ demands for a content, which is cached by other node in its vicinity, i.e., $\Pr \left(\mathcal{I}_\kappa=1\right) = \mathcal{P}_{\rm hit}^{\rm d}-\rho_\kappa$, which corresponds to the parameter $\mathcal{P}_{\rm FDTR}^{v,\kappa}$, i.e.,
\begin{equation}
\label{P_v}
\mathcal{P}_{\rm FDTR}^{v,\kappa} = \Pr \left(\mathcal{I}_\kappa=1\right),
\end{equation}
and the probability $\Pr \left(\mathcal{J}_{\tau, \kappa}=0\right)$ is equivalent to
\begin{equation}
\label{Pr (J=0)}
\Pr \left(\mathcal{J}_{\mu,\kappa}=0\right) = 1-\rho_\kappa.
\end{equation}
Now, the parameter $\mathcal{P}_{\rm FDTR}^{w,\kappa}$ is equivalent to the probability that there is at least one node that demands content $c_\kappa$, hence 
\begin{align}
\label{Formula:P_w}
\mathcal{P}_{\rm FDTR}^{w,\kappa} = &1- \Pr \bigg(  \bigcup_{\tau = 1,\tau \ne \kappa}^{N} \mathcal{J}_{\tau,\kappa}=0  \bigg) \notag \\ \mathop  = \limits^{(a)} & 1- \prod_{\tau=1, \tau \ne \kappa}^{N} \Pr \left( \mathcal{J}_{\tau,\kappa} = 0\right) \notag \\ \mathop  = \limits^{(b)} & 1- \left(1-\rho_\kappa\right)^{N-1},
\end{align}
where (a) follows the fact that the requests at all users are independent from each other and (b) follows directly using the Eq. \eqref{Pr (J=0)}. It is clear that choosing an arbitrary node out of $N$ nodes is equal to $\frac{1}{N}$, which is valid for all modes, hence $\mathcal{P}_{u_\kappa}$ can be defined as 
\begin{equation}
\label{P_uk}
\mathcal{P}_{u_\kappa} = \frac{1}{N}.
\end{equation}
For any arbitrary node, the collaboration probability for all modes can be defined by taking expectation over all possible values for $\kappa$. Now, by substituting the Eqs. \eqref{P_v}, \eqref{Formula:P_w}, and \eqref{P_uk} in Eq. \eqref{P vw}, we can get the final expression as in Eq. \eqref{Formula: Probe FDTR}, i.e.,  
\begin{equation}
\mathcal{P}_{\rm{FDTR}}^{\rm{d}} = \frac{1}{N} \sum_{\kappa=1}^{N}\left(\mathcal{P}_{\rm{hit}}^{\textup{d}}-\rho_\kappa \right)\left(1-\left(1-\rho_\kappa \right)^{N-1}\right).
\end{equation}

\subsection{Proof of Theorem \ref{Theorem:StochasticCaching} }
\label{proof for random caching theorem}
Due to lack of space, we conduct the proof for $\mathcal{P}_{\rm HDTX}^{s}$ yet the methodology for the rest of equations in Theorem \ref{Theorem:StochasticCaching} is the same. To make the proof more feasible, we first consider an example, in which there are three users (i.e., $\text{U} = \{u_1, u_2, u_3\}$) and a library of contents with size $m=4$ ( i.e., $\text{L} =\{ c_1, c_2, c_3, c_4 \}$) and we build a table of outcomes along with the associated permutations matrix $\mathbf{A}$ and column vectors $r_i(\mathbf{A})$, $\mathbf{W}$, $\mathbf{X}$, $\mathbf{Y}$, and $\mathbf{Z}$ as in Table \ref{tab:permut}. We will explain all those matrices in the sequel. As discussed in subsection \ref{subsub:stochasticaching}, there are $m^{N}$ possible outcomes for all users that cache contents at random and these permutations for this specific example are indicated in the columns of $\mathbf{A}$ (i.e., columns $c_1$, $c_2$, and $c_3$ are associated to three users $u_1$, $u_2$, and $u_3$, respectively). The number of rows in this table is $m^N = 4^{3} = 64$, which are associated to the vector $r_i(\mathbf{A})$, $i \in \{1,2,\dots, m^N\}$. Now, we pick up an arbitrary node among $N = 3$ nodes, namely one of users $u_1$, $u_2$, and $u_3$. According to explanations in subsection \ref{subsub:stochasticaching}, since the cached contents as well as the random requests are independent and identically distributed (i.i.d) among all users, there is no preference in choosing an arbitrary node. Thus, without loss of generality, we pick up user $u_1$ that cache content $c_1$. The rest of column vectors, namely $\mathbf{W}$, $\mathbf{X}$, $\mathbf{Y}$, and $\mathbf{Z}$ are formed based on the values given by the permutations matrix $\mathbf{A}$, and we will explain them in the sequel.
\begin{table} [!ht]
	\centering
	\caption{Table of Caching Permutations for N=3 and m=4}
	\begin{tabular}{c|*3c|c|c|c|c}
		\multicolumn{1}{c}{} &\multicolumn{3}{c}{$\mathbf{A}$} & \multicolumn{1}{c}{} & \multicolumn{2}{c}{}   &  \\ 
		$r_i(\mathbf{A})$ & $c_1$ & $c_2$ & $c_3$ & $\mathbf{W}$ & $\mathbf{X}$ & $\mathbf{Y}$ & $\mathbf{Z}$\\
		\hline
		1 & 1 & 1 & 1 &$\rho_2$+$\rho_3$+$\rho_4$ & $\mu_1$ & $\mu_1 \mu_1$ & $ 1-(1-\rho_1)^{2} $\\
		2 & 1 & 1 & 2 & $\rho_3$+$\rho_4$ & $\mu_1$ &$\mu_1 \mu_2$ & $ 1-(1-\rho_1)^{2} $ \\ 
		3 & 1 & 1 & 3 & $\rho_2$+$\rho_4$ & $\mu_1$ &$\mu_1 \mu_3$ & $ 1-(1-\rho_1)^{2} $ \\
		4 & 1 & 1 & 4 & $\rho_2$+$\rho_3$ & $\mu_1$ &$\mu_1 \mu_4$ & $ 1-(1-\rho_1)^{2} $ \\
		\vdots & 1 & \vdots & \vdots & \vdots & $\mu_1$ &\vdots & \vdots \\
		16 & 1 & 4 & 4 & $\rho_2$+$\rho_3$ & $\mu_1$ & $\mu_4 \mu_4$ & $1-(1-\rho_1)^{2} $ \\
		\hline
		17 & 2 & 1 & 1 & $\rho_3$+$\rho_4$ & $\mu_2$ & $\mu_1 \mu_1$ & $ 1-(1-\rho_2)^{2} $\\
		18 & 2 & 1 & 2 & $\rho_3$+$\rho_4$ & $\mu_2$ & $\mu_1 \mu_2$ & $ 1-(1-\rho_2)^{2} $ \\ 
		19 & 2 & 1 & 3 & $\rho_4$ & $\mu_2$ & $\mu_1 \mu_3$ & $ 1-(1-\rho_2)^{2} $ \\
		20 & 2 & 1 & 4 & $\rho_3$ & $\mu_2$ & $\mu_1 \mu_4$ & $ 1-(1-\rho_2)^{2} $ \\
		\vdots & 2 & \vdots & \vdots & \vdots & $\mu_2$ & \vdots & \vdots \\
		\vdots & 2 & 2 & 2 & $\rho_1$+$\rho_3$+$\rho_4$ & $\mu_2$ & $\mu_2 \mu_2$ & $ 1-(1-\rho_2)^{2} $ \\
		\vdots & 2 & \vdots & \vdots & \vdots & \vdots & \vdots & \vdots \\
		32 & 2 & 4 & 4 & $\rho_1$+$\rho_3$ & $\mu_2$ & $\mu_4 \mu_4$ & $ 1-(1-\rho_2)^{2} $ \\
		\hline
		33 & 3 & 1 & 1 & $\rho_2$+$\rho_4$ & $\mu_3$ & $\mu_1 \mu_1$ & $ 1-(1-\rho_3)^{2} $\\
		34 & 3 & 1 & 2 & $\rho_4$ & $\mu_3$ & $\mu_1 \mu_2$ & $ 1-(1-\rho_3)^{2} $ \\ 
		35 & 3 & 1 & 3 & $\rho_2$+$\rho_4$ & $\mu_3$ & $\mu_1 \mu_3$ & $ 1-(1-\rho_3)^{2} $ \\
		36 & 3 & 1 & 4 & $\rho_2$ & $\mu_3$ & $\mu_1 \mu_4$ & $ 1-(1-\rho_3)^{2} $ \\
		\vdots & 3 & \vdots & \vdots & \vdots & $\mu_3$ & \vdots & \vdots \\
		\vdots & 3 & 3 & 3 & $\rho_1$+$\rho_2$+$\rho_4$ & $\mu_3$ & $\mu_3 \mu_3$ & $ 1-(1-\rho_3)^{2} $ \\
		\vdots & 3 & \vdots & \vdots & \vdots & \vdots& \vdots & \vdots \\
		48 & 3 & 4 & 4 & $\rho_1$+$\rho_2$ & $\mu_3$ & $\mu_4 \mu_4$ & $ 1-(1-\rho_3)^{2} $ \\
		\hline
		49 & 4 & 1 & 1 &$\rho_2$+$\rho_3$ & $\mu_4$ & $\mu_1 \mu_1$ & $ 1-(1-\rho_4)^{2} $\\
		50 & 4 & 1 & 2 & $\rho_3$ & $\mu_4$ & $\mu_1 \mu_2$ & $ 1-(1-\rho_4)^{2} $ \\ 
		51 & 4 & 1 & 3 & $\rho_2$ & $\mu_4$ & $ \mu_1 \mu_3$ & $ 1-(1-\rho_4)^{2} $ \\
		52 & 4 & 1 & 4 & $\rho_2$+$\rho_3$ & $\mu_4$ & $\mu_1 \mu_4$ & $ 1-(1-\rho_4)^{2} $ \\
		\vdots & 4 & \vdots & \vdots & \vdots & $\mu_4$ & \vdots & \vdots \\
		64 & 4 & 4 & 4 & $\rho_1$+$\rho_2$+$\rho_3$ & $\mu_4$ & $\mu_4 \mu_4$ & $ 1-(1-\rho_4)^{2} $ \\
	\end{tabular}
	\label{tab:permut}
\end{table}
As can be seen from Table \ref{tab:permut}, it is divided into four equally sized blocks. Each block covers $m^{N-1} = 4^{2} =16$ rows. We first aim to formulate matrix blocking for any size of $\mathbf{A}$. Let us denote $\mathbf{\Omega^{\nu}}$, $\nu \in \{1,2,\dots,m\}$, as the \textit{Block Matrix}, in which $\mathbf{\Omega^{\nu}} \in \mathbf{M}_{m^{N-1} \times N} (\mathbb{N})$, $1 \le \omega_{ij}^{\nu} \le m$, $N \le m$. This block matrix can be obtained by  
\begin{equation}
\mathbf{\Omega^{\nu}} = \mathbf{A} (a_{j\nu}; b_k ),
\end{equation}
where $\nu$ is the content index as described in subsection \ref{subsub:stochasticaching}, $b_k=\{1,2,\dots,N\}$ is the columns index, and $a_{j\nu}$ accounts for the rows index and can be formulated by 
\begin{align}
a_{j\nu}= & \{1+ {\mathbbm{1}_\nu} (\nu-1) m^{N-1},2+ {\mathbbm{1}_\nu} (\nu-1)m^{N-1}, \notag \\& \dots, (1+ {\mathbbm{1}_\nu} (\nu-1))m^{N-1} \} \notag \\ =&  \{ (i+ {\mathbbm{1}_\nu} (\nu-1))m^{N-1} \:|\: i \in \{1,2,\dots,m\} \},  
\end{align}
where ${\mathbbm{1}_\nu}$ is an indicator function and defined by
\begin{equation}
{\mathbbm{1}_\nu } = \left\{ \begin{array}{l}
\begin{array}{*{20}{c}}
0&{;\nu  = 1}
\end{array}\\
\begin{array}{*{20}{c}}
1&{; o.w.}
\end{array}
\end{array}\right.
\end{equation} 
Now, let us explain involving vector columns in Table \ref{tab:permut}. \\
{$\mathbf{W}$}: This column vector indicates possible hitting probability in which the user $u_1$ cannot find its desired content in its vicinity. The reason for this indication comes from the definition of mode $\rm HDTX$ as we discussed in subsection \ref{subsec:UserOperatingModes}. According to definition of this mode, and given any row of permutations $r_i(\mathbf{A})$, user $u_1$ should request any content from the library $\text{L}$ except the contents that appear in the $i^{\text{th}}$ permutation row, e.g., by considering the first permutation $r_1(\mathbf{A})$, namely [1 1 1], user $u_1$ should request any content except $c_1$, which means that it should request one of the contents $c_2$, $c_3$, and $c_4$. The probability of this event is the summation of all probabilities associated to those contents missing in this permutation, namely $\rho_2 + \rho_3 + \rho_4$, which is the first element of column vector $\mathbf{W}$, i.e., $r_1(\mathbf{W})$. Now, let us define a binary random variable $\mathcal{W}_i$ as the event of missing some content indices in the permutation $r_i(\mathbf{A})$, which is given by  
\begin{equation}
\label{Formula:W_i}
{\mathcal{W}_i} = \left\{ \begin{array}{l}
\begin{array}{*{5}{c}}
0&{;{ \textup{ there are missing indices in } r_i(\mathbf{A}) } }
\end{array}\\
\begin{array}{*{5}{c}}
1&{;{  \textup{ all indices are found in } r_i(\mathbf{A}). }}
\end{array}
\end{array} \right.
\end{equation}
Now, the conditional probability of events $\mathcal{W}_i = 0$ and $\mathcal{W}_i = 1$ are respectively defined by
\begin{align}
\label{Formula:PrW0}
\rm Pr \left(\mathcal{W}_i = 0 \:|\: r_i(\mathbf{A}) \right) = & \sum_{\nu = 1}^{m} \rho_\nu - \sum\limits_{\scriptstyle \{\theta_{j}\} \in r_i(\mathbf{A})  \hfill\atop	\scriptstyle \theta_j \neq \theta_k \hfill} {\rho_{\theta_{j}}} \notag \\ = &  1 - \sum\limits_{\scriptstyle \{\theta_{j}\} \in r_i(\mathbf{A})  \hfill\atop	\scriptstyle \theta_j \neq \theta_k \hfill} {\rho_{\theta_{j}}}
\end{align}
\begin{equation}
\label{Formula:PrW1}
\rm Pr \left(\mathcal{W}_i = 1 \:|\: r_i(\mathbf{A}) \right) = 1 - \rm Pr \left(\mathcal{W}_i = 0 \:|\: r_i(\mathbf{A}) \right). 
\end{equation}
One can apply expressions in Eqs. \eqref{Formula:W_i}, \eqref{Formula:PrW0}, and \eqref{Formula:PrW1} for the permutations of block matrix $\mathbf{\Omega^{\nu}}$, which will be used later when we aim to build matrix expressions of the above equations.\\
$\mathbf{X}$: Given user $u_\nu$, this column indicates the probability of the event that this user cached first content of the permutation $r_j(\mathbf{\Omega^{\nu}})$, i.e., the probability of caching content associated to the first column of Table \ref{tab:permut}. We denote this event by $\mathcal{X}_\nu$ and the conditional probability of this event by $\rm Pr (\mathcal{X}_\nu \:|\: r_j(\mathbf{\Omega^{\nu}}))$ which is
\begin{equation}
\label{Formula:PrX}
\rm Pr (\mathcal{X}_\nu \:|\: r_j(\mathbf{\Omega^{\nu}})) = \mu_\nu, 
\end{equation}
$\mathbf{Y}$: Excluding user $u_\nu$ from the permutation $r_j (\mathbf{\Omega^{\nu}})$, the event of caching the rest of contents associated to the rest of the columns of the vector $r_j(\mathbf{\Omega^{\nu}})$ is denoted by $\mathcal{Y}_\nu$ and the conditional probability of this event is denoted by $\rm Pr (\mathcal{Y}_\nu \:|\: r_j(\mathbf{\Omega^{\nu}}))$ and is give by 
\begin{equation}
\label{Formula:PrY}
\rm Pr (\mathcal{Y}_\nu \:|\: r_j(\mathbf{\Omega^{\nu}})) = \prod_{\{\theta_i | \theta_i \neq \nu\} \in r_j(\mathbf{\Omega^{\nu}})}\mu_{\theta_i}. 
\end{equation}
$\mathbf{Z}$: This vector column indicates the probability of the event that there is no user in the vicinity of the user $u_\nu$ that demands for content $c_\nu$. We denote this event by $\mathcal{Z}_\nu$ and the conditional probability of this event by $\rm Pr (\mathcal{Z}_\nu \:|\: r_j(\mathbf{\Omega^{\nu}}))$. The methodology to obtain this probability is similar to the strategy that we have used to obtain Eq. \eqref{Formula:P_w} in the proof of Theorem \ref{Theorem:OptimalCaching}. Hence, we can say
\begin{equation}
\label{Formula:PrZ}
\rm Pr (\mathcal{Z}_\nu \:|\: r_j(\mathbf{\Omega^{\nu}})) = 1-(1-\rho_\nu)^{N-1}.
\end{equation}
All events $\mathcal{W}_i$, $\mathcal{X}_\nu$, $\mathcal{Y}_\nu$, and $\mathcal{Z}_\nu$ are independent from each other and we denote $\phi_{ij\nu}$ as the event of occurrence of all events jointly and the conditional probability of this event as $\rm Pr \left(\phi_{ij\nu}\:|\: r_i(\mathbf{A}), r_j(\mathbf{\Omega^{\nu}}) \right)$, and can be calculated as
%
\begin{align}
\label{Formula:phi_ijv_substituted}
\rm Pr  &\left(\phi_{ij\nu} \:|\: r_i(\mathbf{A}), r_j(\mathbf{\Omega^{\nu}}) \right)=  \rm Pr (\mathcal{W}_i, \mathcal{X}_\nu, \mathcal{Y}_\nu, \mathcal{Z}_\nu \:|\: r_i(\mathbf{A}), r_j(\mathbf{\Omega^{\nu}})) \notag \\& \mathop  = \limits^{(a)}   \rm Pr (\mathcal{W}_i =0 \:|\: r_i(\mathbf{A}))  \rm Pr (\mathcal{X}_\nu, \mathcal{Y}_\nu, \mathcal{Z}_\nu \:|\: r_j(\mathbf{\Omega^{\nu}})) \notag \\& \mathop  = \limits^{(a)}   \rm Pr (\mathcal{W}_i =0 \:|\: r_i(\mathbf{A})) Pr (\mathcal{X}_\nu \:|\: r_j(\mathbf{\Omega^{\nu}})) Pr (\mathcal{Y}_\nu \:|\: r_j(\mathbf{\Omega^{\nu}})) \notag \\& \quad \times Pr (\mathcal{Z}_\nu \:|\: r_j(\mathbf{\Omega^{\nu}})) \notag  \\& \mathop = \limits^{(b)}  \mu_\nu \left( 1 - \sum\limits_{\scriptstyle \{\theta_{j}\} \in r_i(\mathbf{A})  \hfill\atop	\scriptstyle \theta_j \neq \theta_k \hfill} {\rho_{\theta_{j}}}  \right) \times \prod_{\{\theta_j | \theta_j\neq \nu\} \in r_j(\mathbf{\Omega^{\nu}})}\mu_{\theta_j} \notag \\  & \quad \times \left( 1-(1-\rho_\nu)^{N-1} \right),
\end{align}
where (a) follows the independence of all aforementioned events and (b) follows substitution of Eqs. \eqref{Formula:PrW0}, \eqref{Formula:PrX}, \eqref{Formula:PrY}, and \eqref{Formula:PrZ} in the follow-up equation of (a). In our example by substituting values $i=1$, $j=1$, and $\nu =1$, we get the following expression which is associated to the event of permutation $r_1(\mathbf{A}) = $ [1 1 1], i.e.,
\begin{align}
\label{Formula:phi_ijv_substituted_example_single_permut}
\rm Pr \left(\phi_{111}\:|\: \mathcal{W}_1, \mathcal{X}_1, \mathcal{Y}_1, \mathcal{Z}_1\right) & = \mu_1^{3}\big(1- \rho_1\big)\big(1-(1-\rho_1)^{2}\big),
\end{align} 
where $\big(1- \rho_1\big) = \rho_2+\rho_3+\rho_4$ as explained Eq. \eqref{Formula:PrW0}. The expression in Eq. \eqref{Formula:phi_ijv_substituted_example_single_permut} gives the conditional probability of interest for the specific event $\phi_{111}$ among all possible permutations. To get the final expression for the probability of interest $\mathcal{P}_{\rm HDTX}^{s}$ for the HDTX mode, we need to take expectation over all possible outcomes of event $\phi_{ij\nu}$, i.e., 
\begin{align}
\label{Formula:P_HDTX_finalproof}
\mathcal{P}_{\rm HDTX}^{s} = \sum_{i=1}^{m^{N}} \sum_{j=1}^{m} \sum_{\nu = 1}^{N} \rm Pr \left(\phi_{ij\nu}\:|\: \mathcal{W}_i, \mathcal{X}_\nu, \mathcal{Y}_\nu, \mathcal{Z}_\nu\right). 
\end{align}
For our example, the final expression for the HDTX operating mode is
\begin{align}
\mathcal{P}_{\rm HDTX}^{s} = & \mu_1 p_1 [\mu_1^2(\rho_2+\rho_3+ \rho_4)+\dots+\mu_4^2  (\rho_2+\rho_3)]  \notag \\& + \mu_2 p_2 [\mu_1^2(\rho_3+ \rho_4)+\dots+\mu_4^2  (\rho_1+\rho_3)]  \notag \\& +\mu_3 p_3 [\mu_1^2(\rho_2+\rho_4)+\dots+\mu_4^2  (\rho_1+\rho_2)] \notag \\& +\mu_4 p_4 [\mu_1^2(\rho_2+\rho_3)+\dots+\mu_4^2  (\rho_1+\rho_2+\rho_3)],
\end{align}
where, $p_j = (1-(1-\rho_j)^{2})$, $j \in \{1,2,3,4\}$. The above expression can be easily converted to matrix demonstration as in Theorem \ref{Theorem:StochasticCaching}. 


\IEEEQED

\bibliographystyle{IEEEtran}

\bibliography{ref}

\end{document}

%% file: notation.tex
\def\nb0{{\mathbf{0}}}
\def\nb1{{\mathbf{1}}}

\newtheorem{theorem}{Theorem}

\newtheorem{cor}{Corollary}

\def\figref#1{Fig.\,\ref{#1}}%

